# The IDE as a Scriptable Information System (extended version)

Dimitar Asenov
Dept. of Computer Science
ETH Zurich, Switzerland
dimitar.asenov@inf.ethz.ch

Peter Müller
Dept. of Computer Science
ETH Zurich, Switzerland
peter.mueller@inf.ethz.ch

Lukas Vogel
Ergon Informatik AG
Zurich, Switzerland
lukas.vogel@ergon.ch

## ABSTRACT

Software engineering is extremely information-intensive. Every day developers work with source code, version repositories, issue trackers, documentation, web-based and other information resources. However, three key aspects of information work lack good support: (i) combining information from different sources; (ii) flexibly presenting collected information to enable easier comprehension; and (iii) automatically acting on collected information, for example to perform a refactoring. Poor support for these activities makes many common development tasks time-consuming and error-prone. We propose an approach that directly addresses these three issues by integrating a flexible query mechanism into the development environment. Our approach enables diverse ways to process and visualize information and can be extended via scripts. We demonstrate how an implementation of the approach can be used to rapidly write queries that meet a wide range of information needs.

## Keywords

integrated development environments; information system; queries;

## 1. INTRODUCTION

Software development is an information-intense activity. While programming and designing software, developers ask a wide variety of questions [6, 13, 17, 23] and seek information from numerous sources such as the source code itself, compiler output, debug and program analysis tools, version control information, issue tracker, project and API documentation, colleagues, project wiki pages, and community resources like wikipedia.com and stackoverflow.com. In trying to meet their information needs, developers are faced with three issues.

First, developers often need to combine information from more than one source, but tool support for piecing information together is lacking [23]. For example, in order to understand a performance regression, it is useful to combine information from the source code (code structure and control flow), the version control system (recent commits and changes to affected code), performance analysis tools (runtime measurements), and an issue tracker (bugs associated with relevant commits). In situations that require diverse information, developers are forced to manually connect the different pieces of information, which is an error-prone and time-consuming process. Such an information search is also tedious to refine as this usually requires the developer to manually repeat a part of the process.

Second, tools most often present information in a fixed form. Typical presentations include a list of items, a tree-view, or a visual graph. For example, searching with regular expressions results in a list of matches; querying a program with Ferret [3] results in a hierarchical tree-view. These one-size-fits-all presentations are not always a good match for a developer's specific information need (e.g., a visual call graph is better suited for detecting recursion than a hierarchical list), but there is very little or no flexibility for customizing the presentation in existing tools. This could hinder the comprehension of the results and makes domain- and project-specific visualizations impossible.

Third, even after a developer finds the information they need, they often have to take action manually. For example, to understand how a set of methods are called, one has to manually set breakpoints or insert print statements in the code. Another example is a developer manually creating bug reports as a result of an analysis that detects certain code patterns. Repeatedly performing an action manually is time-consuming, error-prone, and frustrating. While some tools support automation (e.g., JunGL [25] and Rascal [7] for refactoring), they cannot integrate arbitrary information sources and are usually limited to certain modifications of source code.

To address these three issues, we designed a query system that integrates directly with developer's primary tool – the integrated development environment (IDE). We make the following contributions:

- An approach for querying information within an IDE, which enables the integration of diverse information resources, flexible result presentations, extensibility via scripts, and automated execution of actions.

- An implementation of the approach in the open-source Envision [2] IDE prototype.

- An evaluation of the applicability of the system in a number of use cases with diverse information needs.

A video demonstrating our system can be seen at youtu.be/kYaRKuUy9rA.

The rest of the paper is organized as follows. In Sec. 2, we motivate our approach with two practical examples. We explain our approach in more technical detail in Sec. 3 and in Sec. 4 we show the wide applicability of our approach using a selection of diverse case studies. In Sec. 5, we provide details of the implementation. We discuss related work in Sec. 6 and conclude in Sec. 7.

## 2. MOTIVATING EXAMPLES

In this section, we will introduce our approach on two practical examples.

## 2.1 Investigating a regression

Suppose that a developer is investigating a recently reported regression, where the incorrect behavior occurs after a specific button is pressed. To investigate this problem the developer will need two main pieces of information: (i) the source code, more specifically, the code that is executed after the handler of the button click, and (ii) the version repository, to see what recent changes could cause this issue.

With current tools, the developer will likely first explore what code is being called from the button handler and manually correlate that to recent changes. This could be a rather time-consuming task if a lot of code is potentially reachable from the handler or if there are many changes that have happened in the mean time. In particularly hard cases, it might pay off to design a specific test case for this regression and run a binary search on the version repository in order to find the offending commit (e.g., using `git bisect`). Both of these approaches are rather time-consuming due to ineffective ways of combining source code information (the call graph) with version information (what changed recently).

Our approach offers an alternative solution. The developer can select the handler of the button in the source code, bring up a query prompt and type:

```
callgraph -nodes | changes -c 5 -nodes
```

The `callgraph` query returns the nodes (methods) in the callee graph of the currently selected method. The bug is likely among these methods, but there may be many of them. To narrow down the search, the methods from the callee graph are piped into the `changes` query, which returns only those methods from the graph that have changed in the last five commits. After the query is executed, the relevant source code fragments will be highlighted and help the developer to more quickly find the issue.

Enabling this workflow are three key components of our approach: **(i)** a **context-sensitive query prompt** that enables developers to quickly type and combine queries; **(ii)** diverse **queries** that can access arbitrary data resources such as the program's source code or version repository; and **(iii)** a **unified data format** that enables queries to be combined in order to refine searches.

## 2.2 Heatmap of code execution

Imagine that a developer wants to get a visual overview of the often-executed parts of the code. The goal is not to optimize specific code, but rather gain a general understanding of which classes are relevant for performance and what is their general function. Thus, it is preferable to see frequently executed methods in a broader context.

Profiling tools typically provide timing information in the form of a chart, graph, or a list. However none of these presentations fits the developer's need in this example, as the code around performance critical methods is also important. The developer will have to manually switch between the profiler and the code they want to explore.

This is one example where the presentation of information is critical for understanding and where our approach's support for flexible visualizations can help. The developer could, for example, export the timings to a CSV file and use

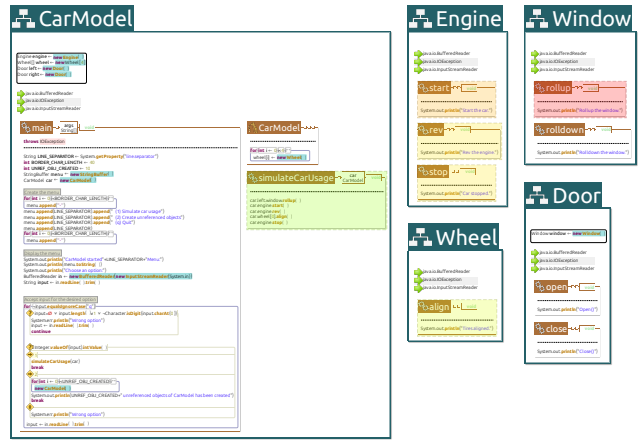

**Figure 1: A heatmap overlayed on top of a visual presentation of code showing a few classes (⊞) and methods (⚙). The heatmap is visualized as a set of translucent overlays on top of methods; each overlay has a color in the spectrum between red and green, indicating how often a method was executed. Similar visualizations are also possible in traditional text-based IDEs, e.g., using line or file highlights.**

a query to import it into the IDE and visualize the results:

```
importProfileCSV profile.csv | heatmap
```

The first time they do this, they will have to write the `importProfileCSV` Python script (about 15-20 lines) that reads the CSV file into the data format understood by our system, but the freely available `csv` Python library makes this task trivial. The read data is piped into the `heatmap` query, which highlights different parts of the code with a color in the red-green spectrum based on the value of a number. An example heatmap is shown in Fig. 1.

This example illustrates two more essential components of our approach: **(iv)** tight integration with a mainstream **scripting language**, allowing easy extension to new information resources and highly-customized queries; and **(v)** **flexible visualizations**, which enable task-specific rendering of information to facilitate comprehension.

## 3. APPROACH

Our goal was to design a system that is highly expressive and extensible by the user in order to satisfy a wide range of information needs. The architecture of our system is shown in Fig. 2. Below, we discuss each component in detail.

## 3.1 Query execution model

The core of our approach is the ability to **compose** and **execute queries**. This functionality is provided by the execution engine, which implements a simple computation model. It enables queries to be connected in a directed acyclic graph, where the edges between queries represent data flow. Such a network of queries is illustrated in Fig. 3.

The execution engine provides each query with a **context**, which is the AST node (e.g., a method) on which the command prompt was invoked. A query has two additional ways

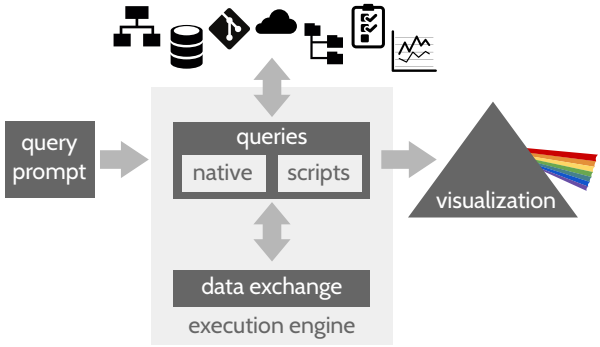

Figure 2: The architecture of our system. Using a query prompt, a developer can invoke and combine queries. Queries can access diverse information resources, make computations, and produce visualizations, and can be either native or implemented via scripts. A unified data exchange format facilitates the cooperation between multiple queries. An execution engine orchestrates query execution and the information flow between queries.[1]

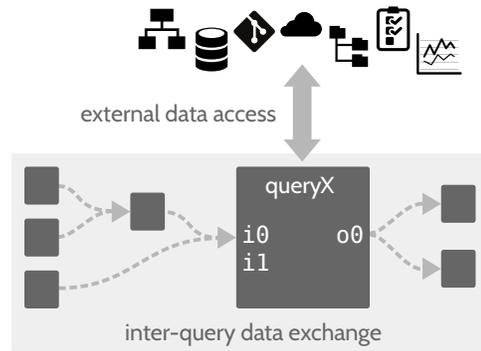

Figure 3: A query network with seven interconnected queries. *queryX* is shown in detail. It has two inputs (`i0`, which is the union of two outputs from other queries and `i1`, which is unused here), and one output (`o0`, which is duplicated).

to access information.

First, a query may be connected to the output of other queries via any number of required or optional inputs, which comprise the **inter-query data exchange**. For example, a query might receive on its input a set of method AST nodes, and output their names as a set of strings. To enable communication between queries, it is key that the data format flowing between queries is unified (see Sec. 3.3).

Second, a query may access (that is, read and modify) **external resources** (see Sec. 3.2.1), such as the program's source code (e.g., to perform a refactoring), call a method of the IDE (e.g., to access the AST or show a message), read from a file (e.g., to import external data), or use a REST service (e.g., to create a bug report).

A query is executed only after all of its inputs are read. When a query is run, it can perform arbitrary computation, which typically includes accessing external resources and computing outputs. Once a query has finished executing, its outputs are forwarded to any downstream queries.

## 3.2 Query types

In principle a query can perform arbitrary computation and use many information resources. However, to facilitate composition, we divide queries into three types (Fig. 4): **resource-access**, **visualization**, and **operator**. Below, we define each type and explain how we designed corresponding queries in order to improve usability.

### 3.2.1 Resource access

Resource-access queries are used to read and modify resources external to the execution engine. They connect a network of queries to the AST and external tools and data. A resource-access query can be a **source of data** for other queries. For example, a query may read the contents of a file or it may extract version information from the project's repository and provide them as inputs to other queries for processing. A resource-access query may also **modify external data**. For example, it could create a new record in a database, or modify the program. There is no limit on the type of external resources a query may access; some common ones are the program code or AST, the version repository, the issue tracker, the file system, and on-line services.

To facilitate composability, we designed resource-access queries according to the following guidelines. First, a resource-access query provides access to only one resource. This restriction allows accessing one resource without imposing requirements on another one. Second, complex resources are accessed by multiple queries, which enables each query to focus on a particular aspect of the resource. For example, when reading the program's AST, one query is used to select nodes while another provides control-flow information. Third, when integrating tools that have a command line interface, we created queries with a similar interface. This enables developers to transfer some of their existing knowledge from the terminal command to the query. For example, in a query that accesses a Git repository, commits can be specified by commit id, branch name, or reference like the `git` command allows.

A noteworthy resource available through resource-access queries is the IDE itself. A query may call available IDE APIs to get information or to perform IDE functions. For example, most IDEs maintain a code model that provides easy access to the program's AST. Such queries are not limited to extracting information. Depending on available IDE APIs, queries might be used for navigation between code fragments, setting breakpoints, running tests, displaying warnings or errors, and refactoring code.

### 3.2.2 Visualization

Visualization queries are used to **render information on the screen**. Different visualization queries can be used to render the same piece of information in different ways in order to better match the specific information needs of the developer. For example, we support three ways to visualize relations between code elements: (i) show the relations using a textual notation – useful for a dense summary; (ii) highlight on screen all code elements that appear in the relations

---
[1]The git logo by Jason Long and icons made by Freepik from www.flaticon.com are licensed by CC BY 3.0.

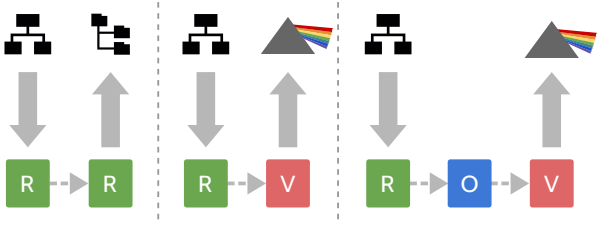

**Figure 4: The three common types of queries: (R) resource-access, (V) visualization, and (O) operator.** On the left a resource-access query extracts information from an AST and forwards it to another query, which writes it to a file. In the middle, this information is being visualized instead of written to file. On the right, an additional filter is inserted to refine what information is displayed.

– useful when searching for particular patterns; (iii) show the relations using arrows between code elements – useful when exploring call graphs or data flow. Some visualizations are provided by the underlying IDE, while others could be done via external tools. Common visualizations in IDEs are highlighting a program fragment, showing a list or a tree of result entries, and displaying error messages and warnings. More visual IDEs could even provide a map of the code that enables intuitive arrow overlays to explore connections between elements or even a heatmap visualization similar to Fig. 1.

Our approach imposes no limit on how information can be visualized. If an IDE exposes general drawing routines, a query could use those to implement an entirely custom visualization. Some IDEs provide a built-in HTML rendering engine, which could be used to easily and quickly implement new ways for visualizing information. Using a combination of HTML5 and Javascript, it is even possible to create interactive visualizations as we demonstrate in Sec. 4.1. This high degree of flexibility is indispensable for domain- and project-specific visualizations.

We designed visualization queries according to the following guidelines in order to make them easier to use. First, each available visualization mechanism has its own query, which makes it clear what will appear on the screen when it is invoked. Second, visualizations impose as few requirements on the input data as possible, so that one visualization can be easily swapped for another. Third, if a visualization query is not explicitly provided by the user, but there is unconsumed data at the end of a query-network execution, a visualization is automatically chosen based on the structure of the result. This frees developers from the need to always explicitly specify a visualization that could be automatically inferred.

### 3.2.3 Operator

Operator queries (operators) are used to **perform internal computation**, for example, to refine results. Operators do not access any external resources but rather help **filter and combine data** in complex query networks. They work solely with the unified data format, which we discuss in Sec. 3.3. As operators work with sets and relations, they naturally map to operations from set and relational algebra such as union, intersect, select, and join. Building on these primitives, we have also pre-defined more elaborate operators in order to simplify common cases. For example, in Sec. 4.2, we demonstrate the `reachable` operator, which filters out elements unreachable from a starting point via a relation – in essence a combination of transitive closure and selection. Such a convenience operator is very useful in answering reachability questions, which are very common [17].

### 3.3 Inter-query data exchange

To enable query composition, all queries communicate using a simple unified structure for exchanging data. This unified exchange structure is a **set of tuples**, because it is sufficiently **expressive** and provides a **simple mental model** for developers to work with.

Each input and output of a query is a set of tuples, and each tuple consists of an arbitrary number of named elements. The names of elements within a single tuple have to be unique. The elements of a tuple may be strings, integers, and references to AST nodes. Each tuple has a tag, which is an identifier that is either explicitly provided or is identical to the name of the tuple's first element. This minimal structure allows us to conveniently encode and access typical structures such as sets, lists, and graphs. For example, the set of all methods whose name starts with `get` could be:

```
{ (node: getAge), (node: getAddress) }
```

The tuple tags could also be provided explicitly:

```
{ node: (node: getAge), node: (node: getAddress) }
```

It is also easy to express relations. For example, all methods transitively called from `rest` could be expressed using the `calls` relation:

```
{ calls: (caller: rest,  callee: watchTV),
  calls: (caller: rest,  callee: sleep),
  calls: (caller: sleep, callee: dream)   }
```

Since there is no restriction on what the set contains, we could also merge the two sets above:

```
{ (node: getAge), (node: getAddress),
  calls: (caller: rest,  callee: watchTV),
  calls: (caller: rest,  callee: sleep),
  calls: (caller: sleep, callee: dream)   }
```

It is also easy to relate information from different data sources. For example, in:

```
{ (commit: "bcdef01", author: "John" ),
  changes: (commit: "bcdef01", node: sleep),
  calls: (caller: rest,  callee: watchTV),
  calls: (caller: rest,  callee: sleep),
  calls: (caller: sleep, callee: dream)   }
```

We can see that John made a change to `sleep`, which is called by `rest`.

Filtering and combining can be easily done based on the name or value of elements, or the tags of tuples. Result visualizations can be automatically selected based on the tags and tuples present in the final output. For example, a tuple with the tag `message` could be shown as a message associated with a code location.

### 3.4 Query prompt

In order to make information access quick and convenient, we designed a specialized input mechanism for invoking queries – the query prompt. Below, we list the key features of this interface.

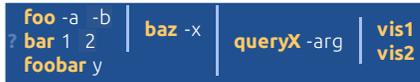

**Figure 5:** A complex composition of queries in the query prompt. The outputs of `foo` and `bar` are merged and forwarded to `baz`, whose output is merged with the output of `foobar` and forwarded to `queryX`. The output of `queryX` is duplicated and forwarded to both `vis1` and `vis2`. The resulting network is the one from Fig. 3.

The prompt is normally hidden and does not take space on screen. Using a keyboard shortcut, the developer can show the prompt on top of an arbitrary code fragment. The prompt is **context-sensitive** — it records the location of the cursor inside the source code at the time the prompt was shown and forwards it to queries. The queries can then use this context (e.g., a class or a method) to implement their behavior.

To use the prompt, developers simply type the queries they want to execute. This **keyboard-based** input allows experienced developers to efficiently invoke queries. Multiple queries can be composed by typing the pipe character '|'. The similarity of this interface to a typical Unix command prompt ensures that developers are already **familiar** with the interaction flow and composition using pipes.

The input field of the prompt is not a standard text box, but a custom widget that allows **non-linear** queries. Fig. 5 shows an example of non-linear input. The prompt does not allow the creation of an arbitrary query network, and uses at most one input and one output per query. Developers can create parallel paths by pressing a keyboard shortcut and can direct data flow using multi-line pipes. There are currently two multi-line pipes: joining and subtracting. Both types have a single output that is duplicated among all downstream queries. A joining multi-line pipe outputs a union of all of its inputs. A subtracting multi-line pipe subtracts from its first input all remaining inputs and outputs the result. Even though this mechanism cannot construct an arbitrary query network, we find that it is sufficiently expressive for many practical tasks.

While we find the query prompt a convenient mechanism, the rest of our system is independent of it and there are other ways for invoking queries. For example, a lighter approach could be to show a text-input on right click and parse linear commands. This could also be extended with support for named pipes. On the other hand, for creating truly custom queries, a graphical editor could be created, which allows manually wiring queries in a complex network.

### 3.5 Extensibility via scripts and native queries

Using our system, the simplest way to access and manipulate information is to compose existing queries directly on the query prompt and get results immediately. Query composition covers a broad range of common needs with minimal effort, but inevitably, developers will need specialized behavior where composition will not be enough. To support custom information processing we provide two extensibility mechanisms.

The first one is to implement new queries as light-weight scripts, in our case using Python. Scripts allow orchestrating complex query flows and provide access to specialized resources, which is made easier by existing libraries for the scripting language. Scripts have access to a limited API: the context, inputs, and outputs of the query they represent, and the program's AST. This API is enough to make scripts versatile while keeping them extremely simple. For example, Fig. 6 shows the complete script for selecting only if statements that have else branches from the input. Implementing a query via a script is as easy as writing the script file. Any available scripts are directly invokable on the query prompt and scripts are seamlessly composable with other queries. The execution engine translates the inter-query data format to and from the native environment and the scripting language's virtual environment.

```
for node in Query.input.tuples('ast'):
  if isinstance(node.ast, IfStatement):
    if node.ast.elseBranch.size() > 0:
      Query.result.add( node )
```

**Figure 6:** A Python script that filters input nodes.

The second extension mechanism is to create new native queries within the host IDE of our system. Native queries give the developer unlimited power to perform specialized computation and allow deep IDE integration — unlike scripts, native queries have access to all IDE APIs. The drawback of this approach is that it is more demanding and time-consuming than writing a script, since developers will have to, essentially, extend the host IDE (e.g., write an Eclipse plug-in).

In practice, native queries, which provide deep IDE integration, and scripts, which access a custom resource, complement each other well, as we show in Sec. 4.

## 4. CASE STUDIES

To analyze the applicability of our approach, we demonstrate its use in a variety of practical programming scenarios. For each case, we motivate its practical relevance, show a possible solution, and discuss how our approach addresses the needs of programmers. All examples can be expressed in our implementation, which is discussed in Sec. 5. A video demonstrating most of the scenarios below can be seen at youtu.be/kYaRKuUy9rA.

### 4.1 Callgraph of selected method

Programmers frequently need to understand control flow and, in particular, follow paths through several method calls [17]. IDEs often have built-in call graph views, and there is a number of research tools that further facilitate the exploration of call graphs [9, 14, 16]. Let's examine a simple scenario:

*What is the callgraph of this method?*

As this is a commonly needed piece of information, we provide a native query:

```
callgraph
```

Because it is **context-sensitive**, this query will automatically return the callee graph of the method that contains the cursor. The query itself just returns a set of tuples and does not produce any visualization. As no visualization is explicitly provided, the execution engine automatically chooses one based on the structure of the result. By default, results that represent relations between AST nodes are visualized as arrows connecting the nodes' representations on screen as shown on Fig. 7(a). Alternatively, the developer could show all result tuples in a table (Fig. 7(b)):

```
callgraph | table
```

or highlight visually the methods that are part of the call graph without showing arrows (Fig. 7(c)):

```
callgraph -nodes
```

Here, the execution engine will detect that the output is a set of AST nodes and will highlight them on the screen. The visualizations from Figs. 7(a) and 7(c) use Envision's visual code presentation, but analogous visualizations also exist in conventional IDEs: line highlights, and arrows between lines. These three visualization options illustrate an important aspect of our approach: the results of queries are decoupled from their visualizations. This decoupling enables **flexible visualizations** that allow the presentation of the results to more accurately match the information need of the developer. Common presentations such as highlights, graphs, arrows, and tables can be readily provided by the IDE. As we show next, developers can also add their own visualizations.

In some projects or domains, specialized visualizations enable better information comprehension. A very convenient way of achieving custom visualizations is to quickly create them by using HTML and existing developer skills. Full-fledged IDEs such as Eclipse often come with a built-in HTML rendering engine, and our approach allows visualization queries to utilize such capabilities. For example, we used the freely available open-source *vis.js* Javascript library and a custom Python script to implement the visualization from Fig. 7(d). With a total of 80 lines of Javascript and Python code we can run the query:

```
callgraph | toHtmlGraph
```

and get an interactive HTML view of a graph that enables users to select and rearrange nodes. The `toHtmlGraph` script converts the tuple set from our system into an HTML page that uses *vis.js*, which in turn provides the rendering and interactions.

### 4.2 Recently changed recursive methods

Sillito et al. [23] identify the lack of support for writing refined queries and combining information as two of the three major gaps in tool support for answering developers' questions. One such question is:

*Which recursive methods have changed recently?*

This question is a refinement of the questions we showed in both Sec. 2.1 and Sec. 4.1. Compared to the former, it adds the requirement of *recursive* methods. Compared to the latter, it adds the need for another information source, the version repository. Our approach enables answering this question directly:

```
callgraph -global
| reachable -self
| changes -c 5 -nodes
```

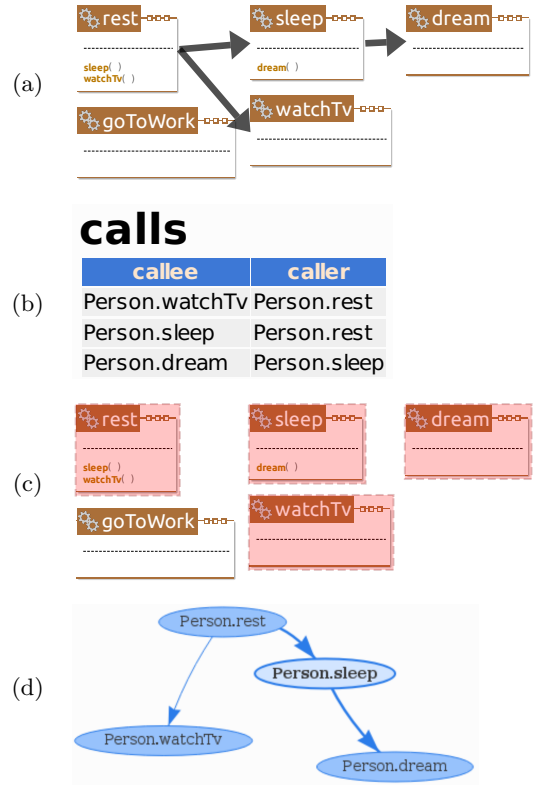

Figure 7: Different ways of visualizing results that describe a relation: (a) arrows between related elements; (b) a table; (c) highlighting relation elements; (d) a custom HTML visualization: in this case, an interactive graph rendered using the `vis.js` Javascript library.

The `-global` argument makes `callgraph` return the call graph of the entire program, ignoring context. The result is then piped into the `reachable` query, which filters AST nodes that cannot reach themselves following the relations from the input. The output contains only recursive methods. Finally, we use the `changes` query to select only those methods from the input that have changed in the last 5 commits.

Two things are noteworthy. First, it was possible to **easily refine** the call graph query by inserting the `reachable` filter in order to get a set of recursive methods. And second, it was also easy to **combine the result with another information source**. Other research tools rarely provide both of these features, and mainstream tools such as regular expression search are thoroughly inadequate for such tasks.

### 4.3 Why is this code the way it is?

Programmers often need to know the reason some code exists or looks the way it does (e.g., questions 8-10 from Fritz and Murphy [6]). There are multiple ways to interpret and answer this question, but a common approach is to connect a piece of code with version control and bug database information. Here is how a query answering this question could look:

```
ast -type Statement -topLevel
| changes -intermediate
| join change.id,commit.message,ast
  -as data
| associatedBugs
```

This is a more complex query, but building it piece by piece is rather straight-forward. We want information about each statement in a method, so we first get all top-level (non-nested) statements using `ast -type Statement -topLevel`. The result is piped into the `changes` query, which yields change information about all commits that modify any of the input statements. The result consists of two different kinds of tuples: the first one associating each node with the id of the commit where it changed, the second associating a commit id with the commit's meta data (e.g., the commit message). Using the `join` query, we merge those two different kinds of tuples into a single kind that relates nodes and commit messages and that we call `data`. We use this name in the `associatedBugs` script (Fig. 8), which is invoked next. This script scans the text of each commit message, looking for references to issue numbers and fetches the corresponding issues' descriptions from the GitHub issue tracker using a REST request. It uses the freely available `GitHub3` Python library to communicate with GitHub. The output of the script is a set of tuples representing `info` messages, which are automatically shown using standard Envision information bubbles next to the corresponding code as shown in Fig. 9. Developers might want to execute this query often. They can conveniently create an alias to it and use this name to call it in the future. The alias may also appear as a subquery in other even more complex queries.

Functionality for explaining the reason for a piece of code is also available in other tools like the `blame` command for version control systems, or the Eclipse annotate feature. However, our approach allows building this information from **elementary blocks** and precise controlling of what is in-

```python
import re
from github3 import GitHub

gh = GitHub()
repo = gh.repository('username', 'repository')

def referencedIssues(commit):
  for issueId in re.findall('#(\d+)', commit):
    yield repo.issue(issueId)

# Build an HTML message from commit and issue data
for data in Query.input.tuples('data'):
  text = '<b>Commit</b><br/>{}<br/>'.format(
      data.message.replace('\n', '<br/>'))
  for issue in referencedIssues(data.message):
    text += '<b>Issue #{}</b><br/>{}<br/>'.format(
        issue.number, issue.title)

  t = Tuple([ ('message', text),
      ('ast', data.ast), ('type', 'info') ])
  Query.result.add(t)
```

Figure 8: A Python script that fetches issue information from a GitHub repository.

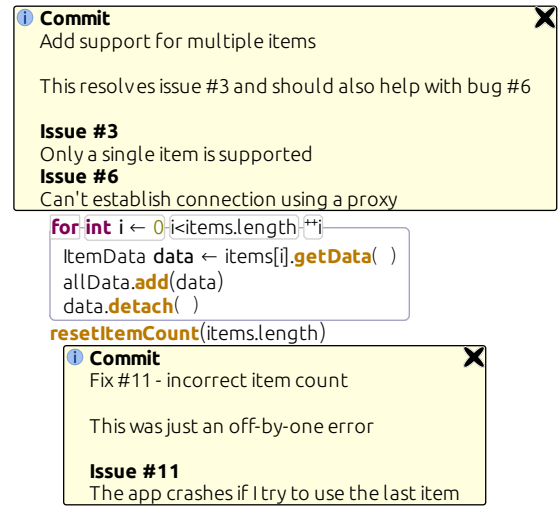

Figure 9: A for-loop and a method call with corresponding explanations for why they were last changed. The information bubbles are standard visualizations in the Envision IDE.

cluded. Developers can also link to arbitrary **additional sources** to fit the answer of this question to their needs.

### 4.4 Which upstream changes possibly conflict with mine?

This question might arise in fast moving projects where developers' local branches may quickly diverge from the development branch. Answering this question precisely is impossible in general, but one way to get an approximate answer is to compare local changes to changes from the remote branch. One possible query to perform this comparison and the corresponding result are illustrated in Fig. 10.

We run two queries in parallel that fetch the latest changes and highlight the changed nodes in different colors. The prompt's support for **non-linear queries** enables the flow of query results to be split and joined to form a more complex query graph, and can often yield a simpler and more intuitive solution compared to a linear approach.

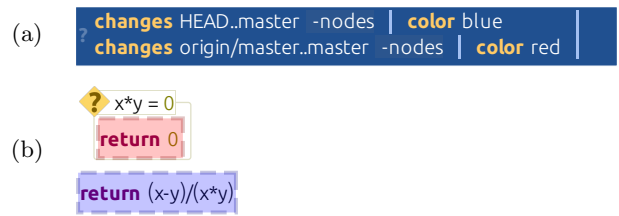

Figure 10: (a) Two parallel queries that will highlight local changes and changes between the master branch and the `origin` repository in different colors. (b) The resulting highlights on part of the code.

## 4.5 Instability metric

A common application of program query tools is to compute software metrics [1, 19]. To illustrate the computation of metrics using our approach, we will compute an instability metric [18]. The instability $I$ of a Java package could be defined as:

$$I = \frac{\textit{Efferent Couplings}}{\textit{Efferent Couplings} + \textit{Afferent Couplings}}$$

Where *Efferent Couplings* is the number of classes inside the package that depend on (import) classes outside the package, and *Afferent Couplings* is the number of classes outside of the package that depend on classes within the package. The higher the instability, the easier it is to change a package without affecting other packages. To compute this metric we could use the following query:

```
ast -type Class -global
| instability
| table
```

This query fetches all `Class` AST nodes, forwards them to the `instability` script, and displays the results in a table. The `instability` script (Fig. 11) iterates over all classes, collecting dependency information in order to compute the metric. Our approach's support for scripts enables the easy **computation of metrics** and more generally of code analyses.

## 4.6 Modifying recursive methods

Programmers often have to change existing code at scale. Development tools typically provide a limited set of refactoring options, and one is lucky if one's use case is covered by these. The fallback solution is most often textual search and replace using regular expressions. However, many situations are non-standard and are simply not expressible using regular expressions. Let's consider one such example – a variation of Sec. 4.2:

*Modify the program so that each recently changed recursive method prints the values of its arguments when called.*

This modification is for instance useful if a program crashes due to a stack overflow after recent changes and one wants to better understand what code is being executed. We could add the necessary print statements using the query:

```
callgraph -global
| reachable -self
| changes -c 5 -nodes
| insertArgPrinting
```

Like we showed in Sec. 4.2, we first get all recursive methods that have changed recently. Then we pass these methods to the `insertArgPrinting` script (Fig. 12), which inserts code that prints the name and lists all arguments of each method.

Three things are worth noting here. First, we executed a query which **modified a resource**, in this case the source code. Second, this modification uses both non-trivial program properties (recursion) and information other than the source code itself (version information). Thus, it is outside the reach of regular expressions and also impossible in typical refactoring languages such as JunGL [25] or Rascal [7], which cannot integrate additional information sources into refactoring decisions. Third, the support for scripts makes it easy to perform program edits within the overall query mechanism. Just like any other query, edits can also depend

```python
# Returns the fully qualified package of a node
def packageOf(node):
  package = ''
  node = node.parent
  while node:
    if type(node) is Module:
      package = node.symbolName() + '.' + package
    node = node.parent
  return package

# Returns a list of all packages a class imports
def dependsOnPackages(aClass):
  result = []
  for decl in aClass.subDeclarations:
    if type(decl) is NameImport:
      package = ''
      name = decl.importedName
      while type(name) is ReferenceExpression:
        package = name.name + '.' + package
        name = name.prefix
      result.append(package)
  return result

allPackages = set()
eff = {}
aff = {}

# Loop over input classes to collect
# package dependencies
for tuple in Query.input.take('ast'):
  p = packageOf(tuple.ast)
  allPackages.add(p)
  deps = dependsOnPackages(tuple.ast)
  if deps:
    eff[p] = 1 + (eff[p] if p in eff else 0)
    for dep in deps:
      aff[dep] = 1 + (aff[p] if p in aff else 0)

# Compute the instability of each package
for p in allPackages:
  e = eff[p] if p in eff else 0
  a = aff[p] if p in aff else 0
  i = str( e/(e+a) ) if e+a > 0 else 1
  t = Tuple([('package', p), ('instability', i)])
  Query.result.add(t)
```

Figure 11: A Python script that computes instability of all packages.

on input from other queries and external data resources, which enables **data-driven changes to the code**.

## 5. IMPLEMENTATION

To validate our approach, we have implemented it as a plug-in to the open-source Envision [2] programming environment. Envision offers a more visually-rich code presentation compared to traditional IDEs and allows us to more easily implement diverse visualization queries. The implementation of our approach and Envision itself are open-source on GitHub: github.com/dimitar-asenov/Envision.

Each native query is implemented as a single C++ class,

```python
for t in Query.input.tuples('ast'):
  m = t.ast
  if type(m) is Method:
    call = 'System.out.println("calling ' +
           m.name + ': "'
    for a in m.arguments:
      call += ' + "' + a.name + '=" + ' + a.name
    call += ')'

    m.beginModification('add print statement')
    nodeExpr=AstModification.buildExpression(call)
    printStmt = Node.createNewNode(
      'ExpressionStatement', None)
    printStmt.expression = nodeExpr
    m.items.prepend(printStmt)
    m.endModification()
```

**Figure 12: A Python script that inserts code to print all arguments at the beginning of a method.**

which has access to our framework's infrastructure as well as all IDE APIs. Additional native queries can be added by creating new query classes in new plug-ins. We use existing IDE features to implement queries that show highlights, arrows, and information messages, and to render HTML. The advanced features of the `changes` command (e.g., version information on a per-AST node basis), are enabled by Envision's own fine-grained version control system.

To enable advanced interactions in the command prompt, we parse the user's input on every keystroke and create a parse tree of the current input. The parse tree is used to create and render parallel queries, and it is mapped to a string representation that enables copying and pasting. This string representation allows users to share queries and is also used when creating aliases to queries.

Queries are executed sequentially by a simple data-flow programming engine, which can be easily extended to parallel execution and streaming of data.

We use *boost.python* to integrate Python scripting. Python scripts have access to the inputs, outputs, and context of a query, as well as Envision's AST model. A few helper features are also available, such as the ability to invoke other queries and to modify the program.

## 6. RELATED WORK

### 6.1 Questions developers ask

A number of recent studies investigate what questions developers ask and how they seek answers to these questions.

Ko et al. [13] observed 17 professional developers in 90-minute sessions and recorded the information they needed and how they acquired it. They present a list of 21 information types and associated questions that developers often asked. The authors suggest that tools should be able to share information and let users transform it as needed, both of which our system directly supports.

Sillito et al. [23] observed 27 developers during two studies and compiled a list of 44 questions that were frequently asked during program evolution. They list three areas for improvement of information seeking tools: (i) support for asking more precise and refined questions, (ii) using context when searching for information and displaying results, and (iii) support for combining information. Our approach directly addresses these three gaps with the ability to pipe queries, to use context for queries and show results in-line with the program, and to combine information from different sources.

In three separate studies LaToza and Myers [17] observed that developers often asked reachability questions and suggest that answering such questions is difficult and time-consuming in large code bases. Reachability questions are about a feasible path through a program, for example in the program control flow or data flow graphs. Our approach directly supports reachability questions.

Fritz and Murphy [6] interviewed 11 professional developers about questions they face frequently and learned 78 questions that span different domains such as the source code, bug database, version history, test cases, etc. Most of these questions require linking different information together, which is supported by our approach.

Sadowski et al. [20] collected data from in-browser code search queries of 27 developers at Google and characterized their search behavior. One of their observations is that developers frequently perform quick searches to navigate code and the authors suggest the integration of search tools directly within the IDE to facilitate quick searches without context switching. Our system enables this workflow.

### 6.2 Tools for seeking information

Researchers have developed a variety of languages and techniques for querying source code [5, 8, 19, 21], which have also been analyzed in comparative studies [1, 4]. Some tools also offer a natural language interface [11, 26]. Such query tools offer powerful and efficient ways to query program properties, but many are restricted to the program source and do not integrate additional information resources. Such query engines could be integrated as a single information resource in our platform, which will enable the combination of their output with additional information. For such an integration, it is important that query results are reified with other entities (e.g., AST nodes) so that they are usable in the rest of the system [5].

A number of tools do allow the combination of different data sources. ABSINTHE [10] was designed specifically to enable queries over different versions of software. More generally, as the basis of the Ferret tool, Alwis and Murphy [3] present a model for integrating information from different sources, which they call Spheres. For example, one sphere could represent source code, while another could capture run-time information such as a call stack. Two spheres can be linked if they contain matching elements, and these matchings have to be predefined by the tool designers. Later, Fritz and Murphy [6] proposed another approach for integrating data from different sources – the *information fragment model*. Unlike Ferret, the information fragment model allows the automatic inference of links between different kinds of information so that it can be easily composed. Our approach is different from these approaches in several ways: (i) the link between different sources does not need to be predefined, but the developer has full control over what information is linked; (ii) the simple tuple set interface of queries allows for the easy addition of a wide range of information resources; (iii) result visualizations are flexible to better match specific information needs.

Recognizing the importance of combining information from different sources, Schiller and Lucia [22] formalize a model for inter-plug-in communication and cooperation within an IDE. They suggest that plug-ins should share data and should allow users to put information from plug-ins together via pipes and filters similarly to the Unix Shell and the Windows Powershell. Our system is also inspired by the Unix shell, but additionally is concerned with visualizations and allows more flexible plug-in and script mixtures. In a similar spirit, Kuhn [15] suggests that IDEs should become open platforms that facilitate the data exchange between plug-ins. In his vision, plug-ins should make all data they compute public and available to other plug-ins for consumption. To share data, he suggests that plug-ins use meta-models to describe the data they produce in a unified system. Our approach also features a unifying component – the tuple set exchange format – which we believe is easier for developers to understand and use to compose complex queries.

### 6.3 Visualization of information

There is a wealth of tools for visualizing information related to software. However these are typically coupled to a specific kind of inquiry. Most tools that provide general query capabilities provide only a single way to view results or only basic visualization flexibility. Common result presentations are list or tree views [3, 5, 6, 8, 11, 26] and graphs [16, 21]. Some tools allow configurable views or advanced interfaces. The prototype implementation for Fritz and Murphy's information fragment model [6] presents the results in a tree view, which allows different projections of the data affecting the hierarchy of objects in the tree. SemmleCode [19] can present results as a list or as a number of predefined chart types. Reacher [16], Stacksplorer [9], and Blaze [14] enable the interactive exploration of call graphs. Our approach is more general and, unlike these other tools, offers flexible and easily extensible visualizations.

### 6.4 Scripting actions and refactoring

Existing program querying tools that go beyond displaying information are typically limited to refactoring code. To support the implementation of complex or project-specific refactorings, researchers have designed scripting languages for refactoring such as JunGL [25] or Rascal [7], which offer powerful capabilities to analyze and transform the source code. Our system also enables complex refactorings via Python scripts and it offers two major advantages: (i) scripts can use external resources in addition to the source code, enabling data-driven refactorings; and (ii) modifying code is just a special-case for our system's general support for automating arbitrary actions.

## 7. CONCLUSION AND FUTURE WORK

We showed an approach to turn an IDE into a powerful and customizable information system. We demonstrated how a **unified data exchange format** enables **query composition** and allows developers to combine **diverse information resources**. The results of such queries can be presented using a **wide range of visualizations** enabling a better fit for a developer's particular information needs – a key in improving comprehension. Our approach can also be used for **automating developer actions** and **data-driven refactorings** of code. We showed that the integration of the Python **scripting language** unlocks great extensibility options, providing a platform for rapidly integrating new information resources or performing complex data manipulation. We were inspired by Unix command shells when designing the **context-sensitive query prompt** – a familiar interface for developers and a convenient entry point for the entire system.

We evaluated our approach using diverse case studies. A user study is future research, which would enable us to further asses the usability of our approach.

Another promising research direction is **interactive queries**. For example, query results that represent code locations and a suitable mechanism to explore them could improve navigation inside the IDE, which takes a significant time for developers [12]. Such navigation could use waypoints, similar to tagSEA [24], which could be used as a guided tutorial that introduces new programmers to a particular part of the system or shows the steps for a routine task. Another use for interactive queries is to integrate a compiler as part of a refactoring query. When performing a non-standard refactoring, sometimes developers first manually change a part of the code, deliberately breaking compilation, and then run the compiler in order to manually traverse and fix each compilation error. Perhaps this process could be more automated using a query that runs the compiler and interactively processes each error. Yet another example for the potential of interactive queries is extracting run-time data. For example, we have done preliminary experiments with setting breakpoints from within queries, pausing queries mid-way in order to allow the program to run, and collecting data during its execution when a breakpoint is hit. This ability to collected run-time data effectively makes this data available as another resource for queries and paves the way for execution-based visualizations and refactorings.

We see the continuing transformation of IDEs into fullfledged information systems as promising and inevitable given the high complexity of today's software. As approaches for integrating data resources like ours evolve, we speculate that standards will emerge. These standards should make it easier to integrate information, breaking down barriers to cooperation between tools and making developers more productive.

## 8. REFERENCES

[1] T. Alves, J. Hage, and P. Rademaker. A comparative study of code query technologies. In *Source Code Analysis and Manipulation (SCAM), 2011 11th IEEE International Working Conference on*, pages 145–154, Sept 2011.

[2] D. Asenov and P. Müller. Envision: A fast and flexible visual code editor with fluid interactions (overview). In *Visual Languages and Human-Centric Computing (VL/HCC), 2014 IEEE Symposium on*, pages 9–12, July 2014.

[3] B. de Alwis and G. Murphy. Answering conceptual queries with Ferret. In *Software Engineering, 2008. ICSE '08. ACM/IEEE 30th International Conference on*, pages 21–30, May 2008.

[4] B. de Alwis, G. Murphy, and M. Robillard. A comparative study of three program exploration tools. In *Program Comprehension, 2007. ICPC '07. 15th IEEE International Conference on*, pages 103–112, June 2007.

[5] C. De Roover, C. Noguera, A. Kellens, and V. Jonckers. The SOUL tool suite for querying

programs in symbiosis with Eclipse. In *Proceedings of the 9th International Conference on Principles and Practice of Programming in Java*, PPPJ '11, pages 71–80, New York, NY, USA, 2011. ACM.

[6] T. Fritz and G. C. Murphy. Using information fragments to answer the questions developers ask. In *Proceedings of the 32nd ACM/IEEE International Conference on Software Engineering - Volume 1*, ICSE '10, pages 175–184, New York, NY, USA, 2010. ACM.

[7] M. Hills, P. Klint, and J. J. Vinju. Scripting a refactoring with Rascal and Eclipse. In *Proceedings of the Fifth Workshop on Refactoring Tools*, WRT '12, pages 40–49, New York, NY, USA, 2012. ACM.

[8] D. Janzen and K. De Volder. Navigating and querying code without getting lost. In *Proceedings of the 2nd International Conference on Aspect-oriented Software Development*, AOSD '03, pages 178–187, New York, NY, USA, 2003. ACM.

[9] T. Karrer, J.-P. Krämer, J. Diehl, B. Hartmann, and J. Borchers. Stacksplorer: Call graph navigation helps increasing code maintenance efficiency. In *Proceedings of the 24th Annual ACM Symposium on User Interface Software and Technology*, UIST '11, pages 217–224, New York, NY, USA, 2011. ACM.

[10] A. Kellens, C. De Roover, C. Noguera, R. Stevens, and V. Jonckers. Reasoning over the evolution of source code using quantified regular path expressions. In *Reverse Engineering (WCRE), 2011 18th Working Conference on*, pages 389–393, Oct 2011.

[11] M. Kimmig, M. Monperrus, and M. Mezini. Querying source code with natural language. In *Proceedings of the 2011 26th IEEE/ACM International Conference on Automated Software Engineering*, ASE '11, pages 376–379, Washington, DC, USA, 2011. IEEE Computer Society.

[12] A. J. Ko, H. Aung, and B. A. Myers. Eliciting design requirements for maintenance-oriented IDEs: a detailed study of corrective and perfective maintenance tasks. In *Proceedings of the 27th international conference on Software engineering*, ICSE '05, pages 126–135, New York, NY, USA, 2005. ACM.

[13] A. J. Ko, R. DeLine, and G. Venolia. Information needs in collocated software development teams. In *Proceedings of the 29th international conference on Software Engineering*, ICSE '07, pages 344–353, Washington, DC, USA, 2007. IEEE Computer Society.

[14] J.-P. Krämer, J. Kurz, T. Karrer, and J. Borchers. Blaze: Supporting two-phased call graph navigation in source code. In *CHI '12 Extended Abstracts on Human Factors in Computing Systems*, CHI EA '12, pages 2195–2200, New York, NY, USA, 2012. ACM.

[15] A. Kuhn. IDEs need become open data platforms (as need languages and VMs). In *Developing Tools as Plug-ins (TOPI), 2012 2nd Workshop on*, pages 31–36, June 2012.

[16] T. LaToza and B. Myers. Visualizing call graphs. In *Visual Languages and Human-Centric Computing (VL/HCC), 2011 IEEE Symposium on*, pages 117–124, Sept 2011.

[17] T. D. LaToza and B. A. Myers. Developers ask reachability questions. In *Proceedings of the 32nd ACM/IEEE International Conference on Software Engineering - Volume 1*, ICSE '10, pages 185–194, New York, NY, USA, 2010. ACM.

[18] R. Martin. Oo design quality metrics. *An analysis of dependencies*, 12:151–170, 1994.

[19] O. Moor, D. Sereni, M. Verbaere, E. Hajiyev, P. Avgustinov, T. Ekman, N. Ongkingco, and J. Tibble. *Generative and Transformational Techniques in Software Engineering II: International Summer School, GTTSE 2007, Braga, Portugal, July 2-7, 2007. Revised Papers*, chapter .QL: Object-Oriented Queries Made Easy, pages 78–133. Springer Berlin Heidelberg, Berlin, Heidelberg, 2008.

[20] C. Sadowski, K. T. Stolee, and S. Elbaum. How developers search for code: A case study. In *Proceedings of the 2015 10th Joint Meeting on Foundations of Software Engineering*, ESEC/FSE 2015, pages 191–201, New York, NY, USA, 2015. ACM.

[21] T. Schafer, M. Eichberg, M. Haupt, and M. Mezini. The SEXTANT software exploration tool. *IEEE Transactions on Software Engineering*, 32(9):753–768, Sept 2006.

[22] T. Schiller and B. Lucia. Playing cupid: The IDE as a matchmaker for plug-ins. In *Developing Tools as Plug-ins (TOPI), 2012 2nd Workshop on*, pages 1–6, June 2012.

[23] J. Sillito, G. Murphy, and K. De Volder. Asking and answering questions during a programming change task. *Software Engineering, IEEE Transactions on*, 34(4):434–451, July 2008.

[24] M.-A. Storey, L.-T. Cheng, I. Bull, and P. Rigby. Waypointing and social tagging to support program navigation. In *CHI '06 Extended Abstracts on Human Factors in Computing Systems*, CHI EA '06, pages 1367–1372, New York, NY, USA, 2006. ACM.

[25] M. Verbaere, R. Ettinger, and O. de Moor. JunGL: A scripting language for refactoring. In *Proceedings of the 28th International Conference on Software Engineering*, ICSE '06, pages 172–181, New York, NY, USA, 2006. ACM.

[26] M. Würsch, G. Ghezzi, G. Reif, and H. C. Gall. Supporting developers with natural language queries. In *Proceedings of the 32nd ACM/IEEE International Conference on Software Engineering - Volume 1*, ICSE '10, pages 165–174, New York, NY, USA, 2010. ACM.